\documentclass[aps,pre,twocolumn,showpacs]{revtex4}
\usepackage[dvips]{graphicx}
\usepackage{epsfig}
\usepackage{amsmath}
\usepackage{amsfonts}
\usepackage{amssymb}
\usepackage{wasysym}

\def\be{\begin{equation}}
\def\ee{\end{equation}}
\def\bfi{\begin{figure}}
\def\efi{\end{figure}}
\def\bea{\begin{eqnarray}}
\def\eea{\end{eqnarray}}
\begin{document}
\voffset=+1truecm
\title{Modelling the brain as an Apollonian network}
\author{Gian Luca Pellegrini$^1$, Lucilla de Arcangelis$^2$, 
Hans J. Herrmann$^3$ and Carla Perrone-Capano$^4$}

\affiliation
{
$^1$ Department of Physical Sciences,
University of Naples Federico II, 80125 Napoli, Italy\\
$^2$ Dept. of Information Engineering and CNISM, 
Second University of Naples, 
81031 Aversa (CE), Italy\\  
$^3$ Computational Physics, IfB, ETH-H\"onggerberg, 
Schafmattstr. 6, 
8093 Z\"urich, Switzerland\\
$^4$ Dept. of Biological Sciences, University of Naples "Federico II", 
80134, Naples, 
Italy and IGB "A.Buzzati Traverso", CNR, 80131 Naples, 
Italy. }

\begin{abstract}

Networks of living neurons exhibit an avalanche mode of activity,
experimentally 
found in organotypic cultures. Moreover, experimental studies of morphology 
indicates
that neurons develop a network of small-world-like  connections, with the 
possibility
of very high connectivity degree. Here we study a recent 
model based on self-organized criticality, which consists in an electrical 
network with threshold firing and 
activity-dependent synapse strengths. We study the model on a 
scale-free network, the Apollonian network, which presents many features of 
neuronal systems. The system exhibits a power law distributed avalanche 
activity. 
The analysis of the power spectra of the electrical signal
 reproduces very robustly the power 
law behaviour with the exponent 0.8, 
experimentally measured in electroencephalograms (EEG) spectra. 
The exponents are found quite stable with respect to initial configurations 
and 
strength of plastic remodelling, indicating that universality holds for a 
wide class of brain models. 

\end{abstract}

\pacs{87.19.La, 05.65.+b, 05.45.Tp, 89.75.-k}

\maketitle 

\section{Introduction}
\label{intro} 

Neuronal networks exhibit diverse patters of activity, including 
oscillations, synchronization and waves. 
During neuronal activity, each neuron can 
receive inputs by thousands of other 
neurons and, when it reaches a threshold,
 redistributes this integrated activity 
back to the neuronal network. Recently 
a neuronal activity based on avalanches has been 
observed in organotypic cultures from coronal slices of rat cortex \cite{beg} 
where neuronal avalanches are
 stable for many hours \cite{beg2}. More precisely, 
recording spontaneous local potentials continuously by
a multielectrode array, has shown that activity initiated at one electrode
might spread to other electrodes not necessarily contiguous, as in a wave-like
propagation. Cortical slices are then found to exhibit a new form of activity,
producing several avalanches per hour of different duration, in which 
non-synchronous
activity is spread over space and time. By analysing the size and duration of
neuronal avalanches, the probability distribution reveals a power law 
behaviour, 
suggesting that the cortical network operates in a critical state. The 
experimental
data indicate for the avalanche size distribution a slope varying between 
-1.2 and -1.9,
depending on the accuracy of the time-binning procedure, with a value -1.5 
for optimal
experimental conditions. Interestingly, the power law behaviour is destroyed
when the excitability of the system is
increased, contrary to what expected since the incidence 
of large avalanches should decrease the power law exponent. 
The distribution then becomes bimodal, i.e. dominated 
either by very
small or very large avalanches as in epileptic tissue.
The power law behaviour is therefore the indication of an optimal 
excitability in
the system spontaneous activity. Moreover the avalanche time duration is also 
found
to follow a power law behaviour as function of the duration time normalised 
by the
binning time with an exponent equal to -2.0 followed by an exponential cutoff.
These results have been interpreted relating spontaneous activity in a 
cortical network
to a critical branching process \cite{zap}, indeed the experimental branching 
parameter is very close to the critical value equal to one, at which 
avalanches at all scales exist.  

On the other hand, the dynamics observed in spontaneous brain activity is very 
similar to self-organized criticality (SOC)
\cite{bak,jen,pacz,pacz2}. 
The term SOC usually refers to a mechanism of slow
 energy accumulation and fast energy redistribution driving the 
system toward a critical state, where the distribution of 
avalanche sizes is a power law obtained without fine tuning: no 
tunable parameter is present in the model. 
The simplicity of the mechanism at the basis of SOC has suggested
 that many physical and biological phenomena characterized by power
laws in the size distribution, represent natural realizations of 
SOC. For instance, SOC has been proposed to model 
earthquakes \cite{Bak,Sornette,lda},
the evolution of biological systems \cite{Sneppen}, solar flare 
occurrence \cite{solar}, fluctuations in confined plasma
\cite{plasma}, snow avalanches\cite{snow} and rain fall \cite{rain}. 

Moreover, power law behaviour is observed in power spectra of different time series 
monitoring neural activities. 
Prominent examples are EEG data which are used by neurologists to discern 
sleep phases, diagnose epilepsy and other seizure disorders 
as well as brain 
damage and disease \cite{gev,buz}. 
Another  example of a physiological function which 
can be monitored by time series analysis is the human gait which is controlled 
by the brain \cite{hau}. For all these time series the power spectrum, i.e. 
the 
square of the amplitude of the Fourier transformation double logarithmically 
plotted against frequency, generally features a power law at least over one 
or two orders of magnitude with exponents between 1 and 0.7. 
Moreover,
experimental results show that the neurotransmitter secretion rate exhibits 
fluctuations with time having power law behaviour \cite{teich} and 
power laws are
observed in fluctuations of extended excitable systems driven by stochastic 
fluctuations \cite{chia}.

On the basis of these observations, recently a model based on SOC ideas and taking into account synaptic 
plasticity in a neural network \cite{noi} has been proposed. 
Plasticity is one of the most astonishing properties of the 
brain, occuring mostly during development and
 learning \cite{alb,hen,abb}, and can be defined as the
 ability to modify the structural and functional properties of synapses, 
properties which are thought to underlie memory and
 learning. Among the postulated mechanisms of synaptic plasticity, 
the activity dependent Hebbian plasticity constitutes the most fully
 developed and influential model of how information is stored in neural circuits
\cite{heb,tsi,coo}.
Within a SOC approach the four
most important ingredients for neuronal activity have been introduced, namely
 threshold firing, neuron refractory period,
 activity-dependent synaptic plasticity and pruning. 

The system consists in an electrical network on a square lattice, 
on which each site 
represents the cell body of a neuron, each bond a synapse. Therefore, 
each site is characterized by a potential and each bond by a conductance. 
Whenever at a given time the 
value of the potential at a site is above a certain threshold, 
approximately equal to $-55mV$ for the real brain, the neuron fires, i.e. 
generates an "action potential", distributing charges to its connected 
neighbours in proportion to the current flowing through each bond. 
After firing, a neuron goes back to the resting potential of $-70mV$ 
and remains inactive during
the refractory period, when it is unable to send or receive information from 
other neurons. This time corresponds for real neurons to the physiological 
time needed to reset ion channels after the transmission of the action 
potential through 
the axon. The conductances, on the other hand, represent Hebbian synapses, 
for which the conjunction of activity at the presynaptic and 
postsynaptic neuron modulates the efficiency of the synapse \cite{coo}. 
To this extent, each time a synapse transmits an action potential between 
active neurons, its strength is increased proportionally to the intensity of 
the transmitted signal, whereas synapses inactive during a neuronal avalanche 
have their strength decreased, as for Hebbian rules. Synapses successively 
weakened may have their strength finally set to zero, i.e. are "pruned", 
eliminating that particular connection between neurons. Pruning implies that,
as activity goes on, the initial regular lattice is transformed, some patterns 
are strengthened and the connectivity of some neurons decreased. 
The system exhibits an avalanche 
activity power law distributed with an exponent close to -1.5, as measured for 
spontaneous activity \cite{beg}. The analysis of the power spectra 
of the electrical signal
 reproduces very robustly the power law behaviour with the exponent 0.8, 
experimentally measured in EEG spectra. 
The same value of both exponents is found considering leaky neurons or 
introducing a small precentage of inhibitory synapses, 
indicating that universality holds for a wide class of 
brain models. 

In real brain neurons are known to be able to develop an extremely high 
number of
connections with other neurons, that is a single cell body may receive inputs 
from even a hundred thousand presynaptic neurons. One of the most 
fascinating questions is how an
ensemble of living neurons self-organizes, developing connections 
to give origin to a highly complex system. The dynamics underlying this 
process should be driven both by the aim of realizing a well connected network 
leading to efficient information transmission, and the energetic cost of 
extablishing very long connections. The morphological characterization of a 
neuronal network grown {\it in vitro} has been studied \cite{shefi} by 
monitoring the development of neurites in an ensemble of few hundred neurons 
from the frontal ganglion of adult locusts. After few days the cultured 
neurons have developed an
elaborated network with hundreds of connections, whose morphology and topology 
has been analysed by mapping it onto a connected graph. The short path length 
and the high clustering coefficient measured indicate that the network belongs 
to the category of small-world  networks \cite{watts}, interpolating between 
regular and random networks. However, the system grown {\it in vitro} 
necessarily lacks some features of {\it in vivo} systems, 
therefore the average node connectivity is found equal only to few units and 
the 
"scale-free" feature \cite{amaral} of many real networks was not recovered.   
Small-world  networks are characterized by an efficient information transmission with
a small number of long range connections. The activity dependent brain model 
\cite{noi}
has been implemented on small world networks, by rewiring a small percentage of the square lattice bonds. Again universal scaling behaviour is recovered
for both the avalanche
distribution and the power spectra. The simple rewiring procedure, however, 
only allows
long range connections leaving the average node connectivity equal to a few 
units, as for {\it in vitro} systems.

In this paper we investigate the behaviour of the activity dependent brain model on
scale-free networks, whose feature are closer to the morphology of neuronal networks 
in living brains. Scale-free networks are indeed characterized by a power law 
distribution of the node connectivity, allowing a high number of connections 
per
neuron. We develop the model on the Apollonian network \cite{apo}, that has 
the property of being simultaneously small-world and scale-free and therefore 
exhibits all characteristics 
found for neuronal networks. Analogously to previous studies, we analyse the 
behaviour of the avalanche size and duration distribution and the power 
spectra related to electrical activity. We also study a system composed by 
both excitatory and inhibitory sysnapses, to be closer to real brains.
The paper is organized as follows: In sect.\ref{net} the scale-free 
Apollonian network is described, whereas in section
\ref{mod} the activity dependent brain model is presented and the results on 
brain
activity are discussed in section \ref{res}. Concluding remarks are given
in section \ref{conclu}.

\section{Apollonian network}
\label{net} 

The Apollonian network has been recently introduced \cite{apo} in a simple 
deterministic
version starting from the problem of space-filling packing of spheres according
to the ancient Greek mathematician Apollonius of Perga. In its classical version
the network associated to the packing gives a triangulation that physically
corresponds to the force network of the packing. 
One starts with the zero-th order triangle of corners
$P_1, P_2, P_3$, places a fourth site $P_4$ in the center of the triangle 
and connects it to the three corners ($n=0$). 
This operation will divide the original triangle in three smaller
ones having in common the central site. The iteration $n=1$ proceeds 
placing one more site in the center of each small triangle and connecting 
it to the corners (Fig.1). At each iteration $n$, going from 0 to $N$,
then the number of sites increases by a factor 3 and the coordination 
of each already existing site by a factor 2. 
More precisely, at generation $N$ there are
$$m(k,N)=3^N, 3^{N-1},3^{N-2},\ldots,3^2,3,1,3$$
vertices with connectivity degree 
$$k(N)=3,3\times 2,3\times 2^2,\ldots,3\times 2^{N-1},3\times 2^N,2^{N+1}+1$$
respectively where the two last values correspond to the site $P_4$ and the 
three corners  $P_1, P_2, P_3$. 
The maximum connectivity value then is the one of the
very central site $P_4$, $k_{max}=3\times2^N$, whereas the sites inserted at 
the $N$-th iteration will have lowest connectivity 3.

\begin{figure}
\includegraphics[width=7cm]{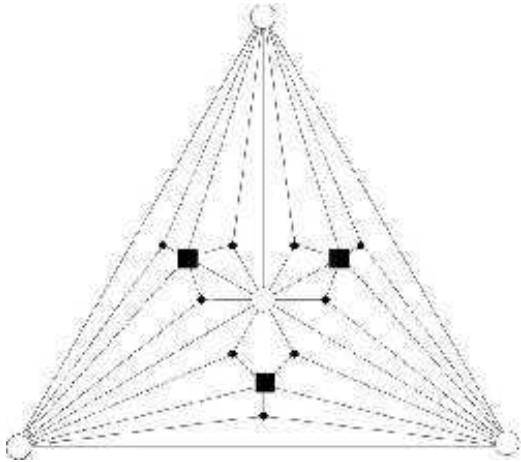}
\caption{
Apollonian network for $N=2$: iterations $n=0,1,2$
are symbols $ \bigcirc , \blacksquare , \bullet$, respectively.
} \label{Fig1}
\end{figure}

The important property of the Apollonian network is that it is scale-free. 
In fact, it has been shown \cite{apo} that the cumulative distribution of 
connectivity degree
$P(k)=\sum_{k'\ge k} m(k,N)/N_N$, where $N_N=3+(3^{(N+1)}-1)/2$ 
is the total number
of sites at generation $N$, has a power law behaviour with $k$. More precisely,
$P(k)\propto k^{1-\gamma}$, with $\gamma =\ln 3 /\ln 2 \sim 1.585$. Moreover 
the network 
has small-world features. This implies \cite{watts} that the average length of 
the shortest
path $l$ behaves as in random networks and grows slower than any positive 
power of $N$,
i.e. $l\propto (\ln N)^{3/4}$. Furthermore the clustering coefficient $C$ is 
very high 
as in regular networks ($C=1$) and contrary to random networks. 
For the Apollonian network $C$ has been found to be equal to $0.828$ in the 
limit of large $N$.
On this basis the Apollonian network appears to have all the features 
typical of 
neuronal networks: small-world property found experimentally \cite{shefi} and 
possibility 
of very high connectivity degree (scale-free). Moreover it also presents bonds 
connecting sites of all lengths. Also this last feature is 
characteristic of neuronal
networks  in brain cortex, where the length of an axon connecting the 
pre-synaptic with the post-synaptic
 neuron can vary over several orders of magnitude, from $\mu m$ to $cm$.

\section{Activity dependent model}
\label{mod}

On a Apollonian network at generation $N$, we assign at each site a neuron 
at potential $v_i$ and at each bond a synapse of conductance $g_{ij}$. 
Whenever at time $t$ the value of the
 potential at a site $i$ is above a certain threshold 
$v_i \geq v_{\rm max}$,  the neuron  
generates an action potential, distributing charges to connected 
neurons in proportion to 
the current flowing through each bond
$$v_j(t+1)=v_j(t)+ v_i(t) {i_{ij}(t)\over\sum_k i_{ik}(t)} \eqno (1)$$
where $v_j(t)$ is the potential at time $t$ of site $j$, connected to 
site $i$, 
$i_{ij}= g_{ij} (v_i-v_j)$ and the sum is extended to all $k$ sites connected to site $i$ that are at a potential $v_k < v_i$. After firing 
a neuron is set to a zero resting potential. 

The conductances can be initially all set equal or else random between 0 and 1, whereas 
the neuron 
potentials are uniformly distributed random numbers between 
$v_{\rm max} - 2$ and $v_{\rm max} - 1$. In agreement with the SOC scenario, 
the initial state for the voltage is not
relevant since the system evolves toward the same critical state regardless 
of the initial condition. 
The potential is fixed to zero at the three sites 1,2,3 where information can 
flow out of the system. 
The external stimulus can be imposed at one input site chosen either 
fixed or at random, this last case modelling more closely spontaneous 
brain electrical activity.

The firing rate of real 
neurons is limited by the refractory period, i.e. the brief period after the 
generation of an action potential during which a second action potential is 
difficult or impossible to elicit. The practical implication of refractory 
periods is that the action potential does not propagate back toward the 
initiation point and therefore is not allowed to reverberate between the cell 
body and the synapse. In the model, once a neuron fires, it
 remains quiescent 
for one time step and is therefore unable to accept charge from firing 
connected neurons. This ingredient indeed turns out to be crucial for a 
controlled 
functioning of the numerical model. In this way an avalanche of charges can 
propagate far from the input through the system. 

As soon as a site is at or above threshold 
$v_{\rm max}$ at a given time $t$, it 
fires according to Eq. (1). Then the conductance of all the bonds, connecting
to active neurons and that have 
carried a current, is increased in the following way
$$g_{ij}(t+1) =g_{ij}(t) +\delta g_{ij} (t) \eqno (2)$$
where $\delta g_{ij}(t)=A \alpha  i_{ij}(t)$, with $\alpha$ being a 
dimensionless parameter 
and $A$ a unit constant bearing the dimension 
of an inverse potential. After applying 
Eq. (2) the time variable of the 
simulation is increased by one unit. 
Eq. (2) describes 
the mechanism of increase of synaptic strength,
tuned by the parameter $\alpha$. This 
parameter then represents the ensemble of 
all possible physiological factors influencing
 synaptic plasticity, many of 
which are not yet fully understood.
 
Once an avalanche of firings
 comes to an end, the 
conductance of all the bonds with non-zero conductance is 
reduced by the
 average conductance increase per bond, 
$\Delta g = \sum_{ij, t} \delta g_{ij} (t)/ N_b$,
where $N_b$ is the number of bonds with non-zero conductance. 
The quantity $\Delta g$
depends on $\alpha$ and on the response of the brain to a given 
stimulus. In this way the network
 "memorizes" the most used paths 
of discharge by increasing their conductance, whereas the less used synapses 
atrophy. Once the conductance of a bond is below an assigned small value 
$\sigma_t$, it is removed, i.e. is set equal to zero, which corresponds to 
what is known 
as pruning. This remodelling of synapses mimicks the fine 
tuning of wiring that occurs 
during "critical periods" 
in the developing brain, when neuronal activity 
can modify the synaptic circuitry, once the basic patterns of brain wiring 
are established \cite{hen}.
These mechanisms correspond to a Hebbian form of activity dependent
 plasticity, where the
 conjunction of activity at the presynaptic and 
postsynaptic neuron modulates the efficiency 
of the synapse \cite{coo}. 
To insure the stable functioning of neural circuits, both strengthening and
 weakening rules of Hebbian synapses are necessary to avoid instabilities 
due to
 positive feedback \cite{des}. However, differently from the
 well known Long Term Potentiation (LTP) and Long Term Depression (LTD)
mechanisms, the modulation of synaptic strength does not depend on
 the frequency of synapse activation \cite{alb,pau,bra}. 
The external driving mechanism to the system is imposed by setting the 
potential of the input site to the value $v_{\rm max}$, corresponding to one 
stimulus. This external stimulus is 
needed to keep the system functioning and therefore mimicks the
 living brain activity. The discharge 
evolves until no further firing occurs, then the next stimulus is applied.

\section{Numerical results}
\label{res}

We consider an Apollonian net at the generation $N=9$ 
(29527 neurons and 177150 synapses). 
The three corner sites of the system have always zero potential and 
represent open boundary conditions. 
The input site is either chosen at random or fixed. 
It is worth noticing that the case of random input sites simulates more 
closely the spontaneous activity of the system. 
Synapses can be excitatory or inhibitory 
with probability $p_{\rm inh}$. Initial conductances can be either all equal to 
$g_0=0.25$ or randomly distributed between 0 and 1. The other parameters in the
simulation are: 
firing threshold $v_{\rm max}=6$ and conductance cut-off for 
pruning  $\sigma_t=0.0001$. Their value does not influence the simulation 
results.

\smallskip 
\noindent
{\it Pruning} 

The strength of the 
parameter $\alpha$, controlling both the increase and decrease of synaptic
strength, determines the plasticity dynamics in the network. In fact, 
the more the system learns strengthening the used synapses, the more the 
unused connections will weaken. We apply a sequence of external stimuli and
we measure the total number of pruned bonds at the end of each avalanche,
$N_{pb}$. This quantity in general could depend on the initial 
conductance $g_0$, therefore the two cases of
all initial conductances equal to 0.25, and uniformly distributed between 0 
and 1, are investigated.

\begin{figure}
\includegraphics[width=8cm]{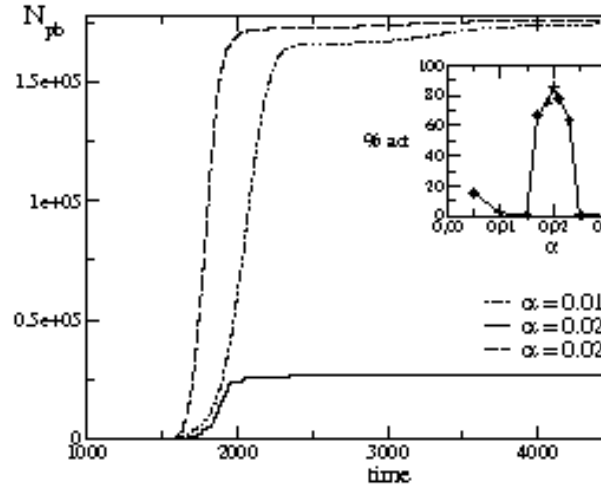}
\caption{Average number of pruned bonds as function of time 
for three different values of  $\alpha$ and equal initial conductances. 
In the inset,
asymptotic number of active bonds as function of $\alpha$. The maximum is for 
$\alpha=0.020$ where are active about $80\%$ of bonds.                                       
} \label{Fig2}
\end{figure}

First the case of equal initial conductances is analysed. For each value of 
$\alpha$ 
the average number of pruned bonds, $N_{pb}$, is monitored as function of time, 
where a time unit corresponds to the application of an external stimulus.
For input sites randomly chosen at each stimulus, Fig.2 shows that 
pruning starts 
after a certain time, since all conductances are initially equal to 0.25, 
and $N_{pb}$ increases more rapidly with time for larger $\alpha$. 
The plateau is reached after 
about 5000 stimuli (for every $\alpha$) after which
$N_{pb}$ increases only of few units in time. From the asymptotic value of 
each curve we can evaluate the asymptotic
number of active bonds as function of $\alpha$ and determine that the value 
of $\alpha$ maximizing the number of active bonds is about 0.020.
This could be interpreted as an optimal value for the
system with respect to plastic adaptation: it maximizes the number of active
connections under the competing strengthening and weakening rules.
                                                                               
In order to understand if pruning acts in the same way on bonds created at
different iterations $n, n=0,\ldots,N$, 
or rather tends to eliminate some particular iteration, the probability 
to prune bonds of different $n$ is evaluated, that is the number of 
pruned bonds over the
total number of bonds for each iteration stage, as function of time. 
Fig.3 shows that the 
plateau is reached at about the same time and the shape of the curve is 
similar for each $n$. However the probability to prune bonds
with large $n$ is higher: These are the bonds created in the last iterations 
and therefore embedded in the interior of the network. This suggests that the 
most active bonds are the long range ones (small $n$), that therefore optimize 
information transport through the network.  In the inset of Fig.3 we show 
the asymptotic number of pruned bonds per generation on a semi-log scale, this 
quantity is well fitted by the exponential behaviour 
$N_{pb}\simeq \exp {n}$. 				 
 
\begin{figure}
\includegraphics[width=8cm]{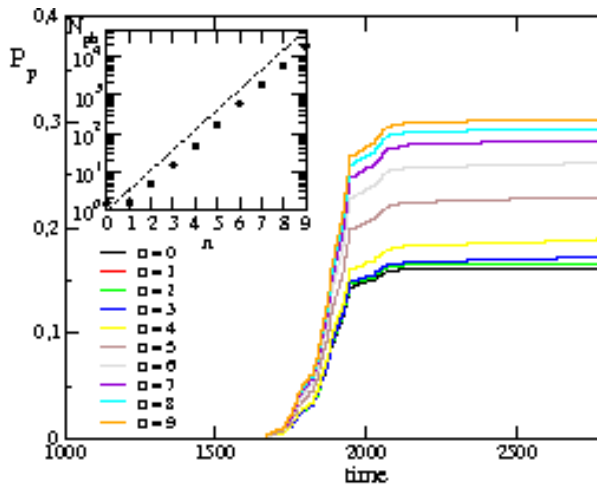}
\caption{(Color online)  Probability of pruning for bonds of different 
iterations $n$ as function of time for equal initial conductances. In the 
inset, the asymptotic $N_{pb}$ (after 5000 stimuli) as function of $n$ with 
the exponential fit $N_{pb}\simeq \exp {1.2n}$.
} \label{Fig3}
\end{figure}

The same analysis has been performed for random initial conductances between 
0 and 1. The results are similar to the previous case. It can be noticed that 
pruning starts already 
at $t=1$, since conductances close to zero are present, and the plateau is 
reached after about 3000 stimuli. The value of $\alpha$ 
which now optimizes the number of active bonds is about 0.030. 
Finally the pruning behaviour for different iterations is similar to the 
previous case, with the pruning probability also increasing with $n$ 
exponentially as $N_{pb}\simeq \exp {n}$.

The effect of pruning on the connectivity degree of the network (Fig.4) has 
also been analysed, i.e. the number of sites with a given connectivity 
degree $k$ as function of $k$ in the initial network and after application 
of a given number of external stimuli. 
In order to identify the different stages in the pruning process, 
the inset of Fig.4 shows the total number of pruned bonds as function of time.
After the application of 
few external stimuli, i.e. for a short plastic training, the distribution 
$n(k)$ shows the same scaling behaviour of the Apollonian network. 
As the pruning process goes on, sites vary their connectivity and new values of $k$
appear. The result is that  the scaling behaviour is progressively lost, as 
well as the scale-free character 
of the network, since there is a generalized decrease of connectivity in the 
network. In the analysis of spontaneous activity it is therefore important 
to impose 
a not too extended plastic training in order to avoid an excessive decrease
of connectivity degree of the network.
 
\begin{figure}
\includegraphics[width=8cm]{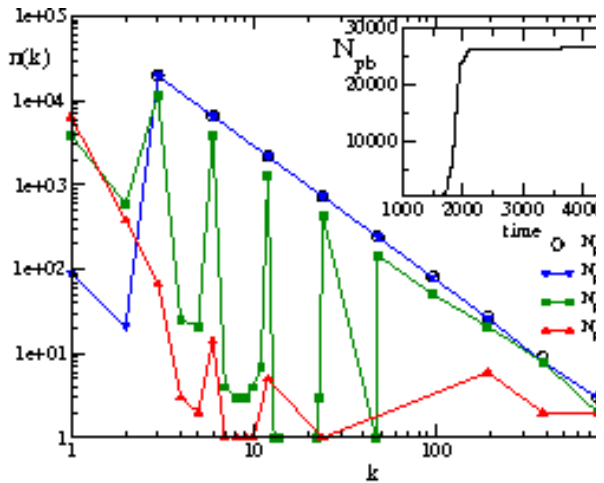}
\caption{(Color online)  Connectivity degree distribution $n(k)$ at
different pruning stages $t$ for equal initial conductances and $\alpha=0.020$. 
In the inset the corresponding behaviour of the number of pruned bonds. 
}
\label{Fig.4}
\end{figure}

\smallskip 
\noindent
{\it Spontaneous activity: avalanche distributions}

After training the system applying plasticity rules during $N_p$ external 
stimuli, we now submit the
system to a new sequence of stimuli with no modification of synapsis
strength. The response of the system to this second sequence models the brain spontaneous activity, which is analysed by measuring the avalanche size 
distribution $n(s)$, 
the time duration distribution $n(T)$, and the power spectrum $S(f)$. 

\begin{figure}
\includegraphics[width=8cm]{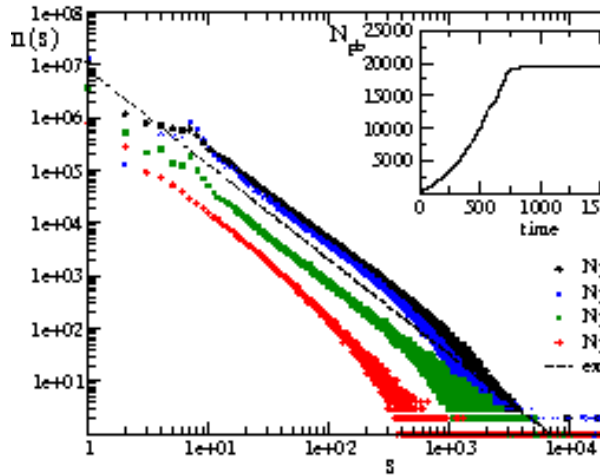}
\caption{(Color online) Avalanche size distribution for different values
of $N_p$, random initial conductances, $\alpha =0.030$ and random input site.
In the inset the corresponding behaviour of the number of pruned bonds.
}
\vskip+1cm
\label{Fig.5}
\end{figure}

The avalanche size distribution $n(s)$ consistently exhibits power law 
behaviour for different values of model parameters. 
Fig.5 shows the avalanche size distribution for different values of $N_p$, 
including also the case $N_p=0$ (no plasticity training) for random initial 
conductances. 
We notice that, for fixed size $s$, increasing $N_p$ decreases the number 
of 
avalanches of that size, suggesting that strong plasticity remodelling 
decreases activity. The exponent appears to be independent of  $N_p$ as long 
as the number of pruned bonds, 
$N_{pb}$, is far from the plateau (see inset in Fig.5). Similar results are 
found for equal initial conductances,
The value of the exponent is $\sigma=1.8\pm 0.2$ and is stable with
respect to variations of the parameter $\alpha$ for both equal and random 
initial conductance. 
This value is  compatible within error bars with the value found in the 
experiments of Beggs and Plentz \cite{beg}, $1.5\pm 0.4$, and with previous 
results of the model 
on both regular and small world lattices. This suggests that the high 
level of connectivity reduces the
probability of very large avalanches but does not change substantially the 
spontaneous
 activity behaviour. For larger $N_p$, the distribution exhibits an 
increase in 
the scaling exponent and finally looses the scaling behaviour for very large 
$N_p$ values in the plateau regime for the number of pruned bonds.

\begin{figure}
\includegraphics[width=8cm]{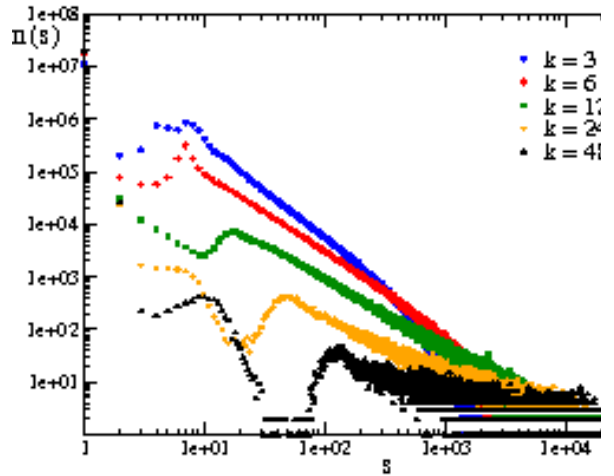}
\caption{(Color online) Avalanche size distribution for input sites randomly
chosen among sites with the same connectivity degree $k$. Only distributions
for small
$k$ are shown, for higher $k$ the scaling behaviour is lost (random initial
conductances, $\alpha=0.030, N_p=100$).
}
\label{Fig.6}
\end{figure}

It is important to investigate the role of the choice of fixed input site,  
since in the Apollonian network, contrary to the regular network, sites may 
have very different connectivity degree.
Fig.6 shows the avalanche size distribution for input sites randomly chosen 
among sites with given connectivity degree $k$.  
In this way it is possible to detect solely the effect due to the connectivity 
of the input site, eliminating all other effects due to the particular 
position of the input 
site in the network. Power law behaviour is found for connectivity degree of 
the input site up to $k=12$. The scaling exponent  decreases with increasing 
connectivity degree $k$ of the input site, that is for larger $k$ larger 
avalanches become more probable. However, if the connectivity degree increases 
further, the scaling behaviour is lost. This is due to the fact that an input 
site with very high connectivity must distribute its charge to many connected 
sites and therefore the network activity will be damped already at the initial 
site. Therefore, to reproduce the experimentally observed scaling behaviour,
 the fixed input site should be chosen with low connectivity 
degree ($k\le 12$). The avalanche size distribution for fixed input site with 
connectivity $k=3$ or $k=6$ exhibits power law behaviour with the same 
exponents found for random input site: $\sigma=1.8\pm 0.2$ for equal  and  
random initial conductances.

At time $t=0$ a neuron is activated by an external stimulus initiating the 
avalanche. This will continue until no neuron is at or above threshold. The 
number of avalanches lasting a time $T$, $n(T)$, as function of $T$ exhibits 
power law behaviour (Fig.7) with an exponential cutoff.
The scaling exponent is found to be $\tau=2.1\pm 0.2$  for equal 
and random initial conductances.  
This value is found to be stable with respect to different $\alpha$ (Fig.7) and 
$N_p$, provided that the number of pruned bonds $N_{pb}$ is lower than 
the plateau for that value of $\alpha$.
Moreover it does not depend on the choice of the input site, either fixed or random.
Finally both values agree within error bars with the value 2.0, exponent 
found experimentally by Beggs and Plentz \cite{beg}. 

\begin{figure}
\includegraphics[width=8cm]{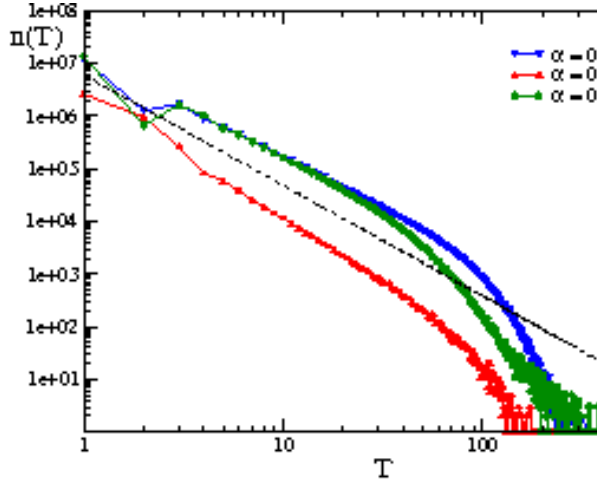}
\caption{(Color online) Avalanche duration distribution for different 
values of $\alpha$ (random initial conductances, random  input sites, 
$N_p =500$).
The dotted line has slope 2.1.} 
\label{Fig.7}
\end{figure}

\smallskip 
\noindent
{\it Power spectra for spontaneous activity}

In order to compare the results of the Apollonian network with EEG medical 
data, the power spectrum of the resulting time series can be calculated. 
For this purpose, the number of active neurons is monitored as function of 
time during spontaneous activity. Fig.8  shows an example of neuronal 
activity where avalanches of 
all sizes can be generated in response to the external stimulus. The power 
spectrum is calculated as the squared amplitude of 
the Fourier transform as 
function of frequency, averaged over many sequences of external stimuli. 

\begin{figure}
\includegraphics[width=8cm]{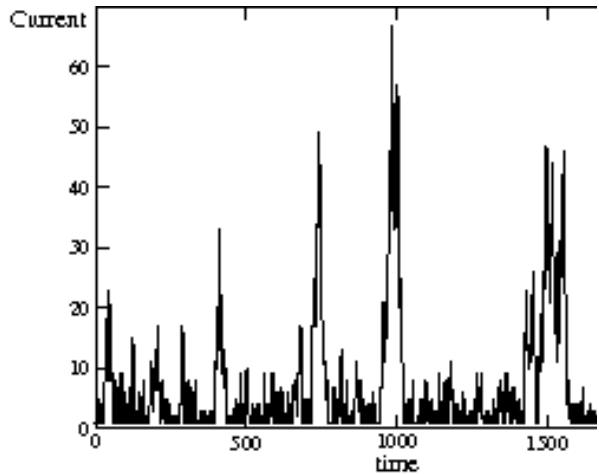}
\caption{Total current flowing in the system as function of time. 
Avalanches of all sizes can be observed.
}
\label{Fig.8}
\end{figure}

Fig.9 shows the spectrum for equal initial conductances and different 
values of $N_p$. 
For $N_p=0$, i.e. when no plasticity mechanism is applied, the spectrum has a 
behaviour $1/f$, characteristic of SOC. For values of $N_p$ different from 
zero but before $N_{pb}$ reaches the plateau, one can distinguish two 
different regimes: a power law behaviour with exponent $\beta=0.8\pm 0.1$ at 
high frequency, followed by a crossover toward white noise at low frequency. 
However, for $N_p=2000$ 
(close  to the plateau value for $N_{pb}$) the scaling behaviour with 
exponent 0.8 is detected over a wider frequency range. 
The difference between $\beta=1$ for $N_p=0$ and $\beta\simeq 0.8$ 
for higher $N_p$, suggests that the existence of plasticity rules reduces the 
power spectrum exponent reaching agreement with experimental EEG 
spectrum \cite{fre,nov}. 
The stability of the exponent with respect to $\alpha$ has also been verified, 
finding  consistently $\beta=0.8\pm 0.1$ at high frequency. 
Finally the power spectrum for fixed input site shows a scaling exponent 
$\beta=0.8\pm 0.1$ over two orders of magnitude. The measured value for 
the power spectra exponent is in agreement with the expected relation
with the scaling exponent of the avalanche duration distribution 
$\beta =3-\tau$, being $-\tau<-1$ \cite{jen}.

The scaling behaviour of the power spectrum can be interpreted in terms
of a stochastic process determined by multiple random inputs \cite{haus}.
In fact, the output signal resulting from different and uncorrelated
superimposed processes is characterized by a power spectrum with power law
behaviour and a crossover to white noise at low frequencies. The crossover
frequency is related to the inverse of the longest characteristic time 
among the superimposed processes. The value of the scaling exponent depends
on the ratio of the relative effect of a process of given frequency on the
output with respect to other processes. $1/f$ noise corresponds to a
superposition of processes of different frequency having all the same relative 
effect on the output signal. In our case the scaling exponent is smaller than
unity, suggesting that processes with high
characteristic frequency are more relevant than processes with low
frequency in the superposition \cite{haus}.

\begin{figure}
\includegraphics[width=8cm]{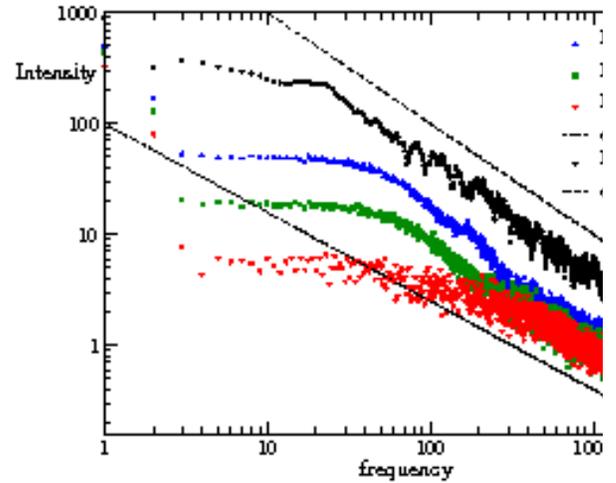}
\caption{(Color online) Power spectra for different $N_p$, equal 
initial conductances, $\alpha=0.020$ and random input sites.
} 
\label{Fig.9}
\end{figure}

\smallskip

\noindent
{\it Inhibitory synapses}

In the mature living brain synapses can be excitatory or inhibitory, namely 
they set the potential of the post-synaptic membrane to a level closer or 
farther, respectively, to the firing threshold. This 
ingredient can be indroduced by considering each synapse inhibitory 
with probability $p_{in}$ and excitatory
 with probability $1-p_{in}$. 
The avalanche size and duration distributions show that the exponents 
$\sigma$ and $\tau$ increase for increasing $p_{in}$, therefore for a high 
percentage of inhibitory
 synapses the probability of large avalanches decreases (Fig.10). 
On the regular lattice for 
$p_{in}=0.5$  no longer power law but exponential behaviour is found 
\cite{noi}. In the present case 
scaling behaviour persists due to the very high connectivity degree,
suggesting that the Apollonian network is better suited to model the neural
connections in the brain.

The power spectra for different values of $p_{in}$ exhibit a complex behaviour.
In fact, for 
a small fraction of inhibitory synapses ($p_{in}\leq 0.05$) the power law 
exponent $\beta$ increases with respect to the case where synapses are all 
excitatory up to a value 1.2. 
Then, for $p_{in}\sim 0.10$, the exponent decreases toward values compatible 
with experimental results, i.e. between 0.7 and 1.0. 
By increasing further the percentage of inhibitory synapses, to values close 
or greater than 0.2, the spectrum becomes the one of white noise.
 
\begin{figure}
\includegraphics[width=8cm]{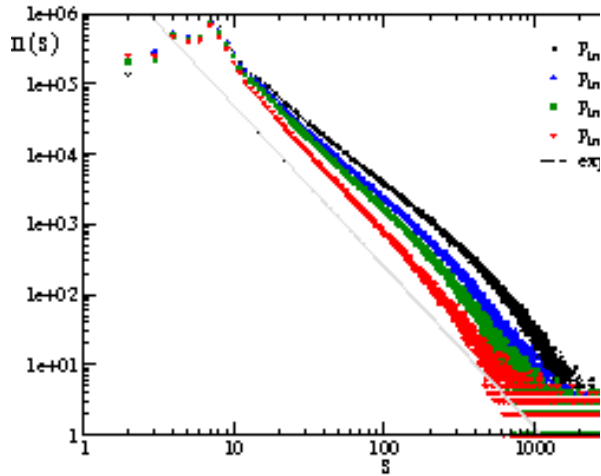}
\caption{(Color online) Avalanche size distributions for different $p_{in}$,
equal initial
 conductances, $\alpha=0.020$, random input sites and $N_p=1700$.
}
\label{Fig.10}
\end{figure}

\section{Conclusions}
\label{conclu}

Extensive simulations have been performed for the activity dependent brain 
model implemented 
on the scale-free Apollonian network. The results are compared with previous 
simulations
on regular and small world lattices and with experimental data. 
We first find the striking result that an optimal value of
of the plasticity strength $\alpha$ exists with respect to
the pruning process. 
Moreover, it appears that synapses of later generations, deeply embedded in 
the network, 
are pruned with higher probability with respect to bonds of the early 
generations, mostly 
long range, that optimize information transmission.
Moreover the avalanche size distribution shows a power law behaviour  
with an exponent 
$\sigma=1.8\pm 0.2$ for equal and random initial conductances. 
This value is compatible with $1.5\pm 0.4$, experimentally found for neuronal 
avalanches and recovered by the model on the square lattice and small world 
networks. 
Also the avalanche duration distribution exhibits power law behaviour with an 
exponential cutoff, in agreement with experimental results of Beggs and 
Plentz \cite{beg}.
 The exponent has value: $\tau=2.1\pm 0.2$ for equal and random initial 
conductance, in agreement with 2.0 found experimentally. 
Furthermore the power spectrum exhibits power law behaviour at high frequency 
with $\beta=0.8\pm 0.1$, in agreement with experimental data \cite{fre,nov}. 
At intermediate frequency the slope becomes greater than unity, crossing
over to white noise at low frequencies.  None of the scaling exponents for 
spontaneous activity in the case of excitatory synapses 
depends on the particular choice for the length or strength of the 
plasticity training and are quite stable with respect to the initial 
conductance configurations. These results suggest that also on 
Apollonian network universal behaviour found for regular and small world 
networks \cite{noi} holds. Furthermore, the scale-free Apollonian net 
provides an excellent description both of the morphology and the electrical
activity properties of the brain.

Acknowledgements. This work was supported by MIUR-PRIN 2004, 
MIUR-FIRB 2001, CRdC-AMRA and EU Network Number MRTN-CT-2003-504712.
H.J.H. acknowledges the Max Planck prize.


\begin{thebibliography}{40}

\bibitem{beg} J. M. Beggs, D. Plenz, J. Neurosci. {\bf 23}, 11167 (2003).

\bibitem{beg2} J. M. Beggs, D. Plenz, J. Neurosci. {\bf 24}, 5216 (2004).
 
\bibitem{zap} S. Zapperi, K.B. Lauritsen, H.E. Stanley, Phys. Rev. Lett. 
{\bf 75}, 4071 (1995).

\bibitem{bak} P. Bak, {\it How nature works. The science of self-organized
criticality}, Springer, New York, 1996.

\bibitem{jen} H.J. Jensen, {\it Self-Organized Criticality}, Cambridge
University Press, Cambridge, 1998.

\bibitem{pacz} S. Maslov, M. Paczuski, P. Bak, Phys. Rev. Lett. {\bf 73},
2162 (1994).

\bibitem{pacz2} J. Davidsen, M. Paczuski, Phys. Rev. E {\bf 66}, 050101(R) 
(2002).

\bibitem{Bak} P.Bak, C.Tang,  J. Geophys. Res. {\bf 94} , 15635 (1989).

\bibitem{Sornette} A. Sornette, 
D.Sornette, Europhys.Lett. {\bf 9}, 197 (1989).

\bibitem{lda} E. Lippiello, L. de Arcangelis, C. Godano, Europhys. Lett.
{\bf 72} , 678 (2005).

\bibitem{Sneppen} P. Bak, K. Sneppen, 
Phys. Rev. Lett. {\bf 71}, 4083 (1993).

\bibitem{solar} E.T. Lu, R.J. Hamilton, 
Astrophys. J. {\bf 380}, L89 (1991).

\bibitem{plasma}  P. A. Politzer, 
Phys. Rev. Lett. {\bf 84}, 1192 (2000).

\bibitem{snow}  J. Faillettaz, F. Louchet, J.R. Grasso, 
Phys. Rev. Lett. {\bf 93}, 208001 (2004).

\bibitem{rain} O. Peters, C. Hertlein and K. Christensen,
Phys. Rev. Lett. {\bf 88}, 018701 (2002)

\bibitem {gev} A. Gevins et al, 
Trends Neurosci. {\bf 18}, 429 (1995).

\bibitem{buz} G. Buzsaki, A. Draguhn, Science {\bf 304}, 1926 (2004). 

\bibitem{hau} J. M. Hausdorff et al., Physica A. {\bf 302}, 138 (2001). 

\bibitem{teich} S.B. Lowen, S.S. Cash, M. Poo , M.C. Teich,
J. Neuroscience, {\bf 17}, 5666 (1997).

\bibitem{chia} D.R. Chialvo, G.A. Cecchi, M.O. Magnasco, Phys. Rev. E {\bf 61},
5654 (2000).

\bibitem{noi} L. de Arcangelis, C. Perrone-Capano, H.J. Herrmann, Phys. Rev. 
Lett. {\bf 96}, 028107 (2006).

\bibitem{alb} T. D. Albraight et al,
Neuron, Review supplement to vol.{\bf 59} (February 2000).


\bibitem{hen} T. K. Hensch, Ann. Rev. Neurosci. {\bf 27}, 549 (2004).

\bibitem{abb} L.F. Abbott, S.B. Nelson, Nature Neurosci. {\bf 3}, 1178 (2000).


\bibitem{heb} D.O. Hebb, {\it The organization of behaviour}, New York: John 
Wiley, 1949.

\bibitem{tsi} J.Z. Tsien, Curr. Opi.n Neurobiol. {\bf 10}, 266 (2000);
G.-Q. Bi, M.-M. Poo, Ann. Rev. Neurosci. {\bf24}, 139 (2001).

\bibitem{coo} S.J. Cooper, Neurosci. Biobehav. Rev. {\bf 28}, 851 (2005).

\bibitem{shefi} O. Shefi, I. Golding, R. Segev, E. Ben-Jacob, A. Ayali,
Phys. Rev. E. {\bf 66}, 021905 (2002).

\bibitem{watts} D. J. Watts, S. H. Strogatz, Nature {\bf 393}, 440 (1998).

\bibitem{amaral} L.A.N. Amaral,A. Scala, M. Barthelemy, E.H. Stanley, Proc. 
Natl. Acad. Sci. U.S.A., {\bf 97}, 149 (2000).

\bibitem{apo} J.S. Andrade, H.J. Herrmann, R.F.S. Andrade, L.R. da Silva, Phys. Rev. Lett. {\bf 94}, 018702 (2005).

\bibitem{des} N.S. Desai, J. Physiol. Paris {\bf 97}, 391 (2003).

\bibitem{pau} O. Paulsen, T. J. Sejnowski, Curr. Opin. Neurobiol. {\bf 10},
172 (2000).

\bibitem{bra} K. H. Braunewell, D.Manahan-Vaughan, Rev. Neurosci. {\bf 12},
121 (2001).

\bibitem{lag} L. F. Lago-Fernandez, R. Huerta,F. Corbacho,J.A. Siguenza,
Phys. Rev. Lett. {\bf 84}, 2758 (2000).

\bibitem{fre}  W. J. Freeman et al, J. Neurosci. Meth. {\bf 95}, 111 (2000). 

\bibitem{nov} E. Novikov, A. Novikov, D. Shannahoff-Khalsa, B. Schwartz,
J. Wright, Phys. Rev. E. {\bf 56}, R2387 (1997). 

\bibitem{haus} J.M. Hausdorff, C.K. Peng, Phys. Rev. E. {\bf 54}, 
2154 (1996).


\end{thebibliography}
\end{document}